\documentstyle[sprocl,epsf]{article}

\def\Journal#1#2#3#4{{#1} {\bf #2}, #3 (#4)}

\def\NPB{{\em Nucl. Phys.} B}
\def\PLB{{\em Phys. Lett.}  B}

\def\NPA{{\em Nucl. Phys.} A}
\def\PR{{\em Phys. Rev.} }

\def\be{\begin{equation}}
\def\ee{\end{equation}}
\def\bea{\begin{eqnarray}}
\def\eea{\end{eqnarray}}

\def\peta{\frac{g^2_A}{2 f^2}} 
\def\ddq{{{\rm d}^Dq \over (2\pi)^D}\,} 
\def\ddl{{{\rm d}^Dl \over (2\pi)^D}\,}
\def\ddqm{{{\rm d}^{D-1}{\bf q} \over (2\pi)^{D-1}}\,}
\def\ddlm{{{\rm d}^{D-1}{\bf l} \over (2\pi)^{D-1}}\,}

\def\dtqm{{{\rm d}^{3}{\bf q} \over (2\pi)^{3}}\,}
\def\dtlm{{{\rm d}^{3}{\bf l} \over (2\pi)^{3}}\,}
\def\dtkm{{{\rm d}^{3}{\bf k} \over (2\pi)^{3}}\,}
\def\bfq{{\bf q}} 
\def\bfk{{\bf k}} 
\def\bfp{{\bf p}}
\def\bfl{{\bf l}}
\def\bfpp{{\bf p '}} 
\def\darr#1{\raise1.5ex\hbox{$\leftrightarrow$}\mkern-16.5mu #1}
\def\){\right)} 
\def\({\left(} 
\def\]{\right]} 
\def\[{\left[}
\def\lskip{\vspace{\baselineskip}}

\newcommand{\eqn}[1]{\label{eq:#1}}
\newcommand{\refeq}[1]{(\ref{eq:#1})}

\newcommand{\Eq}{Eq.~\refeq} 

\newcommand{\beq}{\begin{eqnarray}}
\newcommand{\eeq}{\end{eqnarray}}

\newcommand{\mcal}[1]{{\mathcal #1}}

\newcommand{\makefig}[4]{\begin{figure}[t] 
                          \centerline{\epsfysize=#3 in \epsfbox{#2}} 
                           \caption{\protect{#4} \label{#1}} 
                         \end{figure}}

\newcommand{\maketwofigs}[6]{\begin{figure}[t]
                           \centerline{\epsfysize=#5 in \epsfbox{#2}
                                \hspace{#3 in} \epsfysize=#5 in \epsfbox{#4}}
                                \caption{#6 \label{#1}} 
                                \end{figure}}

\begin{document}

\title{NUCLEON-NUCLEON SCATTERING IN EFFECTIVE FIELD THEORY\footnote{NT@UW-99-34} }

\author{\underline{G. RUPAK}, N. SHORESH}

\address{ Department of Physics, University of Washington, \\
Seattle, WA 98195,USA}


\maketitle\abstracts{I outline the effective field theory (EFT) calculation of
nucleon-nucleon scattering which was recently carried out to
next-to-next-to-leading order (NNLO) by Noam Shoresh and myself. In
this calculation only potential pion contributions are included.
These are the leading contributions from pions. A toy model was
also considered in which potential pions are the only pion effects. The effective range expansion (ERE), which is valid for low energies, is used to derive matching conditions on some of the EFT couplings. Renormalization group flow analysis is used to fix the rest of the couplings leaving no free parameters at NNLO. }

\section{Introduction}
\label{INTRO}

Effective Field Theory (EFT) is useful for systems with a clear
separation of scales. This is the case in nucleon-nucleon scattering, where the light pions have a mass of $m_\pi\sim140 \mbox{MeV}$, and the next particle to
consider, the $\rho$ meson, only comes in at $m_\rho\sim770
\mbox{MeV}$. In describing the two-nucleon system, however, special
care is needed because of the appearance of a new low energy scale,
namely the singlet channel scattering length. The Kaplan-Savage-Wise (KSW)
power counting scheme~\cite{ksw1} was formulated to address this issue.

In the EFT where the pions are included explicitly, the contribution
of loops containing pions can be separated into (a) potential pions and
(b) radiation and soft pions. Potential pions, corresponding to instantaneous pion exchanges, arise when the nucleons in the loop are put on-shell. Since
both nucleons can be put simultaneously on-shell these contributions
are enhanced. The radiation and soft pions enter when the
\emph{pions} are put on shell and describe retardation pion effects.

In this talk I describe the contribution of the potential pions
alone~\cite{gautam}. It is argued below, Subsection~\ref{Nijmegen}, that this is sufficient for a next-to-next-leading order (NNLO) calculation of nucleon-nucleon ${}^1S_0$ scattering amplitude at external energies close to the pion production threshold. At energies much below this threshold, there could be contributions from radiation and soft pions. Nevertheless, the calculation with only potential pions serves as a probe for a
better understanding of the KSW power counting, the appropriate fitting procedure and the role of renormalization group flow analysis. It is worthwhile
to consider a model in which only potential pions exist, and one does
not have to worry about the effects of the unaccounted for radiation
and soft pions. Such is the two-Yukawa model described in
Section~\ref{Toy}. The effective Lagrangian with potential pions for center of mass momentum $p\leq m_\rho/2$ is described in Section~\ref{Lagrangian}. KSW power counting is briefly reviewed in Section~\ref{KSW}. The next-to-leading order (NLO) and NNLO scattering amplitude is calculated in Section~\ref{NLO} and ~\ref{NNLO} respectively. In Subsection~\ref{Matching}, some of the EFT couplings are determined from low energy matching conditions for the amplitudes. Renormalization group (RG) flow analysis in Subsection~\ref{RG} fixes the rest of the couplings. Results for the toy model are presented in Subsection~\ref{toy_results}, whereas in Subsection~\ref{Nijmegen} I have included the corresponding results for real ${}^1S_0$ scattering.

 There are also relativistic corrections to the amplitude at NNLO due to the finite mass of the nucleons. These effects are very small compared to the potential pion contributions and can be neglected.

\section{The Toy Model}
\label{Toy}

 We consider a toy model in which non-relativistic nucleons interact via two instantaneous Yukawa potentials:

\beq
V(r)=-g_\pi\frac{e^{-m_\pi r}}{4\pi r}-g_\rho\frac{e^{-m_\rho r}}{4\pi r}
\eqn{V}
\eeq
($r$ is the distance between the nucleons).
This model is complex enough in the sense that the interactions involve two well separated scales. The Yukawa potentials describe the instantaneous exchange of the $\pi$ and a scalar $\rho$ meson (without the complications of retardation effects~\cite{mehen2}).

The realistic ${}^1S_0$ one pion exchange amplitude
(here given in momentum space) is
\beq
\(-\frac{g_{A}^{2}}{2 f_{\pi}^{2}}\)\frac{p^2}{p^2+m_{\pi}^{2}}=-\frac{g_{A}^{2}}{2 f_{\pi}^{2}}+\frac{g_{A}^{2}}{2 f_{\pi}^{2}}\frac{m_{\pi}^{2}}{p^2+m_{\pi}^{2}},
\eqn{OPEP}
\eeq
where the first term is a contact interaction, and the second is the Yukawa part. 
The contact interaction is not included in the toy model. $g_\pi$ is chosen to reproduce the Yukawa part.
Comparing \Eq{OPEP} and \Eq{V}, we set
\beq
g_{\pi}=\frac{g_{A}^2 m_\pi^2}{2 f_{\pi}^2}.
\eeq

The nucleon mass is taken to be $M=940\mathrm{\ MeV}$, as well as
$m_\pi=140\mathrm{\ MeV}$, $m_\rho=770\mathrm{\ MeV}$, $f_\pi=132\mathrm{\ MeV}$ and
$g_A=1.25$.
$g_\rho$ is tuned to give a large scattering length ($g_\rho=13.5\rightarrow a=-24.7\mathrm{\ fm}$) by numerically solving the Schroedinger equation with this potential. 

The potential pion contribution to the field theory amplitude in this toy model and the real ${}^1S_0$ nucleon-nucleon scattering are the same. One might wonder about the contact piece in Eq.~\refeq{OPEP} that was neglected. As will be explained in Section~\ref{NLO}, in the field theory amplitude the contact piece of the one-pion-exchange (OPE) can be absorbed in the definition of one of the 4-nucleon couplings.

\section{The Effective Lagrangian}
\label{Lagrangian}

The effective Lagrangian for the toy model appropriate for center of mass energies much smaller than $m_\rho$ is presented here. The effects of the short range force, $\rho$ exchanges, are reproduced in the EFT by contact interactions. It is convenient~\cite{ksw2} to separate the effective Lagrangian in terms of the number of nucleon fields present: 
\beq {\cal L} = {\cal L}_1 + {\cal L}_2 + ...,\nonumber \eeq 
where ${\cal L}_n$ describes the propagation or scattering of
  $n$ nucleons. $\mcal{L}_1$ is the usual
 kinetic energy term
\beq \eqn{chiral_Lagrangian}
{\cal L}_1 = N^\dagger (i \partial_0 + \vec{\nabla}^2/2 M)N,
\eeq 
where the isodoublet field N represents the nucleons:
\beq N =\left( \begin{array}{c}
                p\\
                n
              \end{array}\right).
\nonumber  
\eeq 
Next, we write ${\cal L}_{2}^{({}^1S_0)}$, the part of ${\cal L}_2$ 
that has non-vanishing matrix elements between ${}^1S_0$ states. 
 The matrices $P_k$~\cite{ksw2} are used to project onto the ${}^1S_0$
 state,  
\beq P_k \equiv \frac{1}{\sqrt{8}}\sigma_2\otimes\tau_2\tau_k,
\nonumber 
\eeq
where the $\sigma$ matrices act on the nucleon spin space and the
$\tau$ matrices act on the nucleon isospin space. The result is~\cite{ksw1,gautam,ksw2}:  

\beq
\eqn{L_2} 
{\cal L}_{2}^{({}^1S_0)} &=& -{\(\frac{\mu}{2}\)}^{4-D}(C_0 + D_2 m_\pi^2 + D_4
m_\pi^4+\cdots)(N^T P_i N)^\dagger(N^T P_i N)\nonumber\\ 
&{ }& +{\(\frac{\mu}{2}\)}^{4-D}(\frac{1}{8}C_2+ \frac{1}{8}D_4^{(2)}
  m_\pi^2+\cdots)[(N^T P_i N)^\dagger(N^T
P_i(\stackrel{\leftrightarrow}{\nabla})^2 N)+h.c]\nonumber\\ 
&{ }& + \cdots
\eeq
(summation over the repeated isospin index $i$ is implied.)

We use dimensional regularization where the Lagrangian is given in $D$ space-time 
dimensions and $\mu /2$ is an arbitrary mass scale introduced so that the
couplings $C_{2n}$, $D_{2n}^{(2m)}$ have the same units in any dimension $D$. $C_{2n}$ are the coefficients of contact interactions
containing $2n$ derivatives and
 $D_{2n}^{(2m)}$ are coefficients of contact interactions involving 
 $2m$ derivatives and $2(n-m)$ powers of the pion mass. We use  
   the convention $D_{2n}\equiv D_{2n}^{(0)},\ C_{2n}\equiv D_{2n}^{(2n)}$. The ellipses indicate
operators with higher powers of derivatives or pion mass insertions.


The interaction with the pions is described by a non-local term in the action:
\beq
S_{\pi}&=&\int d^D x\  d^D y\(N^T(x)P_i N(x)\)^\dagger g_{\pi}\frac{e^{-m_\pi|{\bf x}-{\bf y}|}}{4\pi|{\bf x}-{\bf y}|}\delta(x^0-y^0)\(N^T(y)P_i N(y)\) \nonumber\\
&=&\int d^D p\  d^D p'\(N^T(p)P_i N(p)\)^\dagger \peta\frac{m_{\pi}^{2}}{{|{\bf p}-{\bf p'}|}^2+m_{\pi}^{2}}\(N^T(p')P_i N(p')\). 
\eeq

Once the effective Lagrangian has been described, one needs a power counting scheme, i.e. a set of rules that determine the relative sizes of diagrams contributing to the scattering process.

\section{The power counting}
\label{KSW}
KSW power counting is motivated by the large scattering length $a$. It can be better understood by looking at the singlet channel amplitude at very low energy where even the pions can be treated as heavy particles and hence integrated out. The effective field theory amplitude is related to the phase shift by 

\beq
{\cal A}=\frac{4\pi}{M}\frac{1}{p\cot\delta-ip},
\eqn{pcot2A} 
\eeq
where $p=\sqrt{M E}$ is the center of mass momentum and $\delta$ is the singlet channel phase shift. For low energies one can write~\cite{bethe} the effective range expansion (ERE): 

\beq  
p\cot\delta&=&-\frac{1}{a}+\frac{1}{2}\sum_{n=0}^{\infty}r_{n}p^{2(n+1)}\nonumber\\
&=&-\gamma+\frac{1}{2}\sum_{n=0}^{\infty}s_{n}\(p^2+\gamma^2\)^{n+1}.
\eqn{ERE2}
\eeq

Here the scattering length $a$ can be arbitrarily large, while the other coefficients $r_0$, $r_1$, ... are assumed to be of natural sizes, $r_n\sim1/\Lambda^{2n+1}$, where $\Lambda\sim m_\pi$ is the cut-off for this low energy theory. $p$ is taken to be small compared to the high energy cut-off $\Lambda$ but not compared to $1/a\ll\Lambda$. The coefficients in the two different ERE expansions are related perturbatively. The first two terms are
\beq
\frac{1}{a}&=&\gamma -\frac{1}{2} {{\gamma }^2} {s_0}+\mcal{O}({{\gamma }^4})\nonumber\\
{r_0}&=&{s_0}+2\gamma^2s_1
    +O({{\gamma }^4}).
\eeq 
The difference between $\gamma$ and $1/a$ enters at NLO, whereas the difference between $s_0$ and $r_0$ is only important at orders higher than NNLO.

Substituting Eq.~\refeq{ERE2} in Eq.~\refeq{pcot2A}, it can be seen that the amplitude has a pole at $p=i\gamma$ and this sets the radius of convergence for the Taylor expansion about $p=0$. Formally identifying $\gamma$, $p\sim Q$, where $Q/\Lambda$ is the expansion parameter, the amplitude Eq.~\refeq{pcot2A} can be expanded about the pole in powers of $p/\Lambda$ and $\gamma/\Lambda$,

\beq
\mcal{A}&=&-\frac{4\pi}{M}\frac{1}{\gamma+ip}\[1+\frac{{s_0}}{2}
  \frac{p^2+\gamma^2}{\gamma +ip}
   +{{\(\frac{{s_0}}{2}\)}^2}
    \frac{{{({p^2}+{{\gamma }^2})}^2}}
     {{{(\gamma +ip)}^2}} \right.\nonumber\\
& &\hspace{1.6in}\left.+\(\frac{{s_0}}{2}
  \)^3\frac{{{({p^2}+{{\gamma }^2})}^
        3}}{{{(\gamma +ip)}^3}}
    +\frac{{s_1}}{2}
     \frac{{{({p^2}+{{\gamma }^2})}^2}}
      {\gamma +ip} +\cdots \]\nonumber\\
&=&-\frac{4\pi}{M}\frac{1}{\gamma+ip}\[1+\frac{{s_0}}{2}
   (\gamma -i p) + 
   {{\(\frac{{s_0}}{2}\)}^2}
   {{(\gamma -ip)}^2} \right.\nonumber\\
& &\left.\hspace{1.4in}+ \(\frac{{s_0}}{2}
  \)^3{{(\gamma -ip)}^3}
     +\frac{{s_1}}{2}
      {{(\gamma -ip)}^2}
       (\gamma  + ip)+\cdots \]\nonumber\\
&\equiv&\sum\limits_{n=-1}^{\infty} \mcal{A}_n,\hspace{.5in} \mcal{A}_n\sim Q^n.\eqn{EREinA}\eeq 

In KSW power counting the field theory amplitude is organized as an expansion in $Q$, Eq.~\refeq{EREinA}. It has been shown~\cite{weinberg} that the leading term in expansion \Eq{EREinA} is obtained by summing up the bubble chains in Fig. \ref{LOfig} with insertions of $C_0$, giving the leading order (LO) amplitude:
\makefig{LOfig}{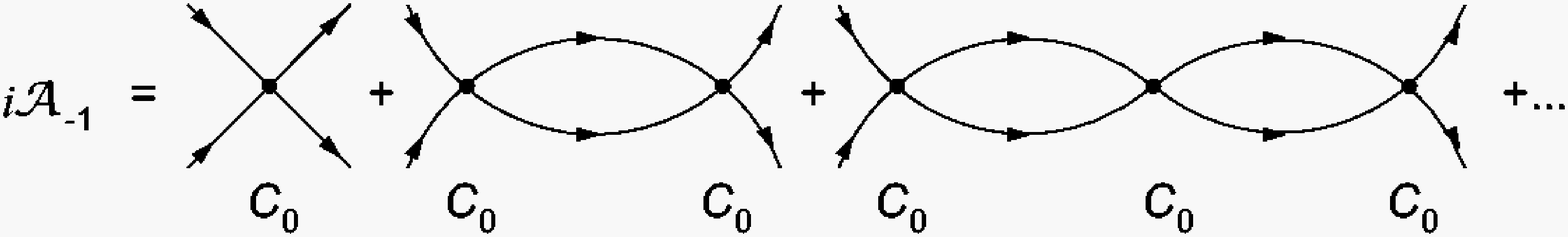}{.8}{Leading order contribution $\mcal{A}_{-1}$ from non-perturbative insertions of $C_0$.}
\beq
  \mcal{A}_{-1}= \frac{-C_0}{1 + i C_0 L},
   \eeq where $L$ is the loop integral
\beq \eqn{L}
L &=& {\( \frac \mu 2\)}^{\rm 4-D}\int\ddq \frac i {\frac E 2 -
q_0 -\frac {q^2} {2 M} + i\epsilon} \frac i {\frac E 2 + q_0
-\frac {q^2} {2 M} + i\epsilon}\nonumber\\ 
&=& -iM {\(\frac \mu
2\)}^{\rm 4-D}\int\ddqm \frac 1 { q^2 - E M
-i\epsilon}=-i\frac M{4 \pi}(X_{sub} + i p).
\eeq

A subtraction scheme that reproduces the KSW power counting is the power divergence subtraction (PDS) scheme that was introduced in~\cite{ksw1}. In PDS one subtracts the poles of $L$ in both $D=4$ and $D=3$
dimension (which gives $X_{sub}=\mu$ in \Eq{L}). Setting
$\mu\sim p$, all the diagrams contributing to the bubble chains are
rendered equal in size ($\sim 1/p$). This justifies summing up the
bubble graphs at LO.

In the theory with pions, pion effects are assumed to be perturbative.
From the PDS scheme follows a power counting
scheme - KSW - in which:

\begin{itemize}
\item{ The expansion parameter is $Q/\Lambda$, where $p,\mu,m_\pi\sim Q$.}
\item{ $\mcal{A}=\sum_{n=-1}^{\infty}\mcal{A}_n,\ \mcal{A}_n\sim Q^n$.}
\item{ The renormalized coupling $C_0=C_0(\mu)$ scales as
    ${1/\mu}\sim 1/Q$, and more generally the renormalized couplings $C_{2n}$, $D_{2n}^{(2m)}$
    scale as $ 1/{Q^{n+1}}$.}
\end{itemize}
Some immediate consequences are:
\begin{itemize}
\item{ The loop integral $L$ scales like $p$ or $\mu$, i.e. ${\cal
      O}(Q)$. Therefore, adding a $C_0$ and a loop $L$ to a diagram
    does not change its order since the powers of $Q$ cancel. It
    is now required to dress each diagram in the theory by $C_0$ to all
    orders. (This is diagrammatically represented by the ``blob'', Fig.~\ref{blobfig}).

\makefig{blobfig}{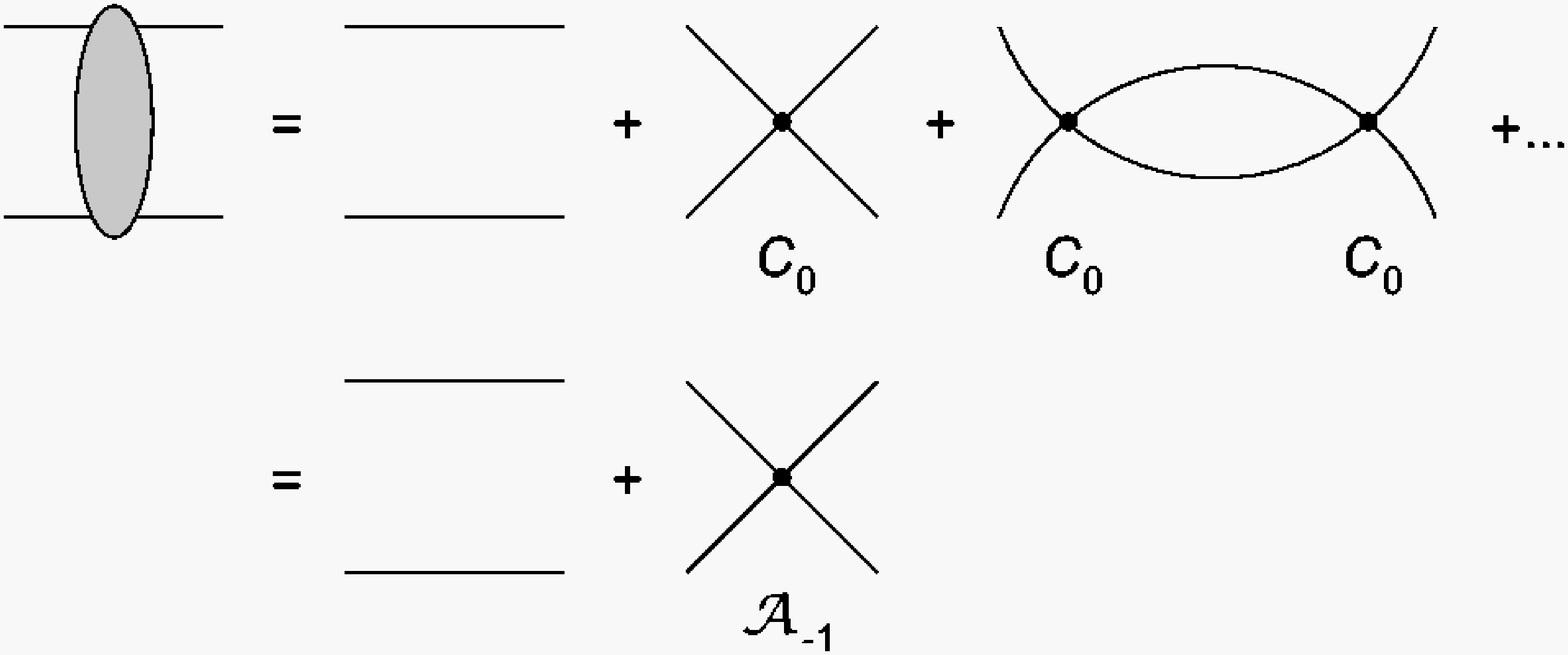}{2}{Definition of the ``blob''.} }
\item{ When the nucleons are put on shell, nucleon propagators scale
    as $1/Q^2$ and loop integrals $\int d q^4$ scale as $Q^5$.}
\end{itemize}
 
This theory should correctly describe the physics at momentum below
$\Lambda\sim m_\rho/2$, which is the scale associated with the cut in the amplitude due to $\rho$ exchange. A calculation to order $Q^n$ is expected to be accurate up
to an error of $\sim(Q/\Lambda)^{n+1}$.
\lskip

It is a key feature of the KSW power counting that the LO amplitude has the right pole, or the correct scattering length. However, at higher orders there are contributions that move the position of the pole in the amplitude and counterterms must be introduced to cancel these effects. This is done by expanding the EFT couplings in powers of $Q$~\cite{gautam,mehen,chen1}:
\beq
C_{0}&=&C_{0,-1}+C_{0,0}+C_{0,1}+\cdots \nonumber\\
C_{2}&=&C_{2,-2}+C_{2,-1}+\cdots \nonumber\\
C_{4}&=&C_{4,-3}+C_{4,-2}+\cdots \nonumber\\
\vdots& &
 \eeq
where the second subscript denotes the scaling with powers of $Q$. 

In the last three sections, all the components of the computational framework of the EFT for nucleons have been identified: the effective Lagrangian (Feynman rules), the regularization and subtraction methods, and the power counting scheme that organizes the diagrams of the theory in a perturbative series. In the following sections we apply this formalism to the calculation of the ${}^1S_0$ scattering amplitude.

\section{The Amplitude at NLO}
\label{NLO}

 In the theory with truly dynamical pions, the contact piece of the pion exchange can be re-absorbed in the definition of $C_{0,0}$. Hence, there is no difference in the potential pion amplitude for this toy model and the real N-N scattering amplitude. The potential pion contribution at NLO has already been calculated before~\cite{ksw1}. I include the calculation here for completeness and clarity.

At NLO, we need to consider local operators that make up the following four-nucleon vertices, classified here according to their Q counting:\vspace{.1 in}

\indent\hspace{0.5in}\begin{tabular}{lcl}
${\cal O}(Q^{-1})$&:&$ -C_{0,-1}$\\[.1 in]
${\cal O}(Q^0)$&:&$-\frac{(\bfq_1^2+ \bfq_2^2)}{2}C_{2,-2}\ ; -C_{0,0}\ ;-m^2_\pi D_{2,-2}$\vspace{.1 in}
\end{tabular}

\noindent with $\bfq_1$, $\bfq_2$ the single nucleon incoming and outgoing momenta, in the center of mass coordinates. Only the linear combination $C_{0,0} + m^2_\pi D_{2,-2}$ appears in the amplitude, and it is denoted by  $B_{0,0}$.
From this point on we adopt the convention that the second subscript is omitted from leading couplings, so that, for instance, $C_0$ stands for $C_{0,-1}$, $C_2$ is really $C_{2,-2}$, etc. 

The diagrams contributing to the NLO amplitude are shown in Fig.~\ref{NLOfig}. 
The amplitude is
\beq \mcal{A}_0= \mcal{A}_0^{(0 \pi)} + \mcal{A}_0^{(\pi)} \nonumber \eeq  
 where $\mcal{A}^{(0 \pi)}_0$ includes all the diagrams with no pion exchange, and $\mcal{A}^{(\pi)}_0$ contains all the diagrams with a single pion exchange.
It is convenient to express Fig.~\ref{NLOfig}(b) and Fig.~\ref{NLOfig}(c) in terms of the loop integrals $P_1$ and $P_2$, so that
\makefig{NLOfig}{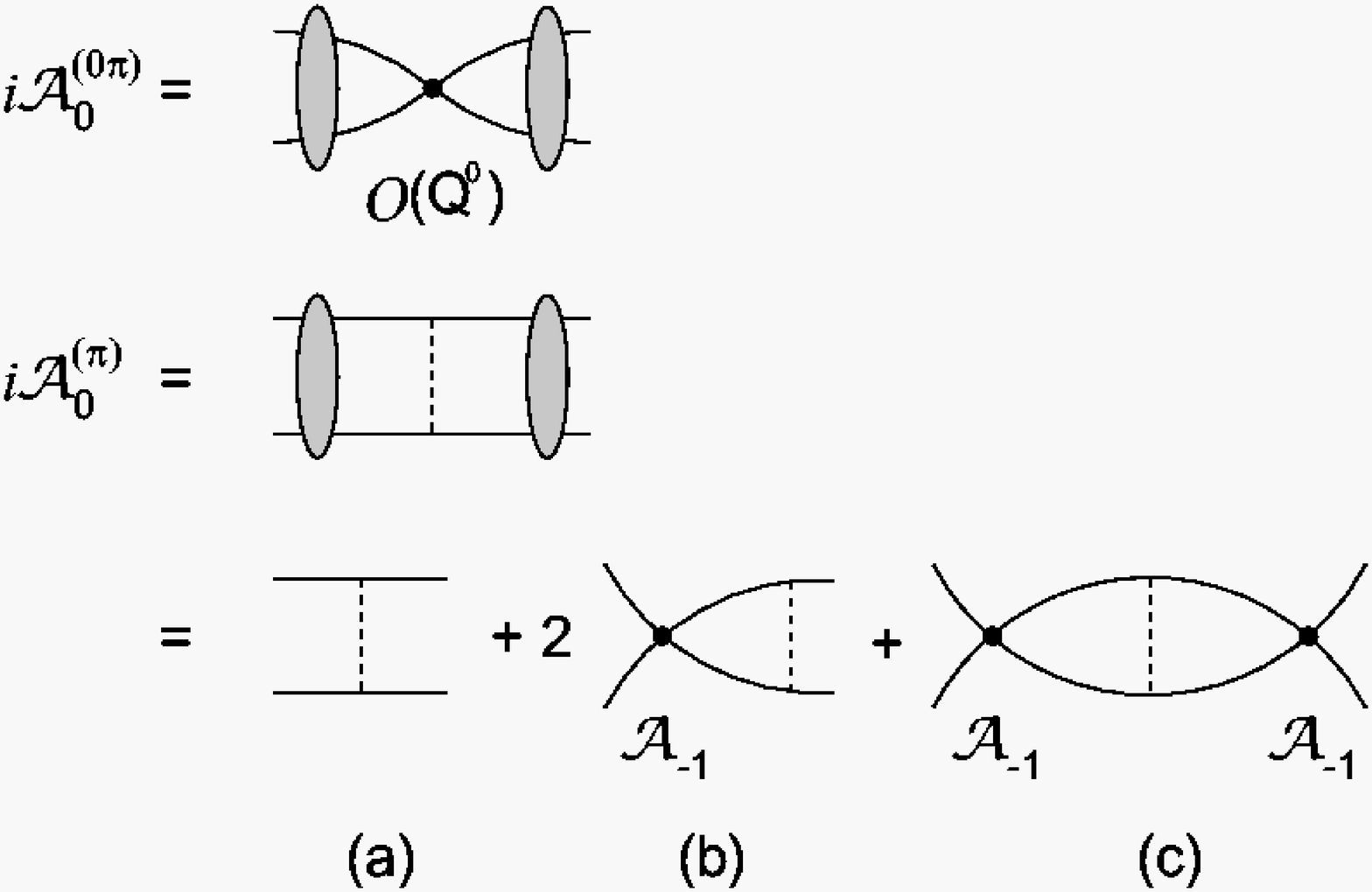}{3}{Next-to-leading order contribution $\mcal{A}_{0}$. }
\beq 
\mcal{A}_0^{(0 \pi)}&=&  {(1+ i\mcal{A}_{-1} L)}^2( - B_{0,0} - C_2 p^2)\nonumber\\
\mcal{A}_0^{(\pi)}&=& \frac{g_A^2}{2 f^2}\frac{m_\pi^2}{4 p^2} \ln (1+\frac{4
  p^2}{m_\pi^2})+ 2\ (\mcal{A}_{-1}) P_1 +i (\mcal{A}_{-1})^2 P_2, 
\eeq 
with the definitions
\beq \eqn{P1}
P_1 &\equiv&
i\peta m^2_\pi\(\frac{\mu}{2}\)^{4-D}\int\ddq\frac i {\frac E 2 + q_0 -\frac{q^2}{2 M}
+ i\epsilon}\nonumber\\
& &\hspace{1.8 in}\times\frac i {\frac E 2 - q_0 -\frac{q^2}{2 M} +
i\epsilon}\frac 1 {({\bf q -p})^2 + m^2_\pi} \nonumber\\ 
&\stackrel{PDS}\rightarrow&\frac 1 {8 \pi} \peta \frac{m^2_\pi
M}{p}\( \arctan(2 \frac p {m_\pi} ) + \frac i 2 \ln (1 +
4\frac{p^2}{m^2_\pi})\) \eeq
and
\beq \eqn{P2}
P_2
&\equiv& i\peta m^2_\pi\(\frac{\mu}{2}\)^{8-2D}\int\ddq\ddl\frac i {\frac E 2 +
q_0 -\frac{q^2}{2 M} + i\epsilon}\nonumber\\
&\times&\frac i {\frac E 2 - q_0
-\frac{q^2}{2 M} + i\epsilon}\frac i {\frac E 2 + l_0
-\frac{{ l}^2}{2 M} + i\epsilon}\frac i {\frac E 2 - l_0
-\frac{{ l}^2}{2 M} + i\epsilon}\frac 1 {({\bf l -q})^2 +
m^2_\pi} \nonumber\\[.1 in]
&\stackrel{PDS}\rightarrow&-i\peta \frac{m^2_\pi M^2}{32 \pi^2}
\left(-\gamma_E + \ln(\pi)+1 \right.\nonumber\\
& &\hspace{1.5 in}\left.- 2\ln\(\frac{m_\pi}{\mu}\) - 2\ln\(1 -
i2\frac p {m_\pi}\) \). \eeq

\Eq{P2}, where only the poles have been subtracted, differs from an earlier calculation of $P_2$~\cite{ksw1} that involved finite subtractions.

\section{The Amplitude at NNLO}
\label{NNLO}

At this order, there are insertions of three kinds of local
operators, listed here
 according to their $Q$ counting:\vspace{.1 in}

\indent\hspace{0.5in}\begin{tabular}{lcl}
${\cal O}(Q^{-1})$&$:$&$ -C_0$\\[.1 in]
${\cal O}(Q^0)$&$:$&$-B_{0,0}\  ; -\frac{(\bfq_1^2+ \bfq_2^2)}{2} C_2$\\[.1 in]
${\cal O}(Q^1)$&$:$&$-B_{0,1}\  ; -\frac{(\bfq_1^2+ \bfq_2^2)}{2} B_{2,-1}\  ; -p^4 C_4$\vspace{.1 in}
\end{tabular}

\noindent There are 3 new couplings that enter at this order: $B_{0,1}\equiv C_{0,1}+m^2_\pi D_{2,-1} +m^4_\pi D_4$,  $B_{2,-1}\equiv C_{2,-1}+ m^2_\pi D_{4}^{(2)}$ and $C_4$. The NNLO amplitude is
\beq \mcal{A}_1=\mcal{A}^{(0 \pi)}_1
+ \mcal{A}^{(\pi)}_1 + \mcal{A}^{(\pi\pi)}_1. 
\nonumber \eeq

\subsection{$\mcal{A}^{(0 \pi)}_1$}
\label{zeropion}

The diagrams in Fig.~\ref{NNLO0pfig} are generated by two insertions of ${\cal O}(Q^0)$ operators
or a single insertion of ${\cal O}(Q^1)$ operator, and give: 

\makefig{NNLO0pfig}{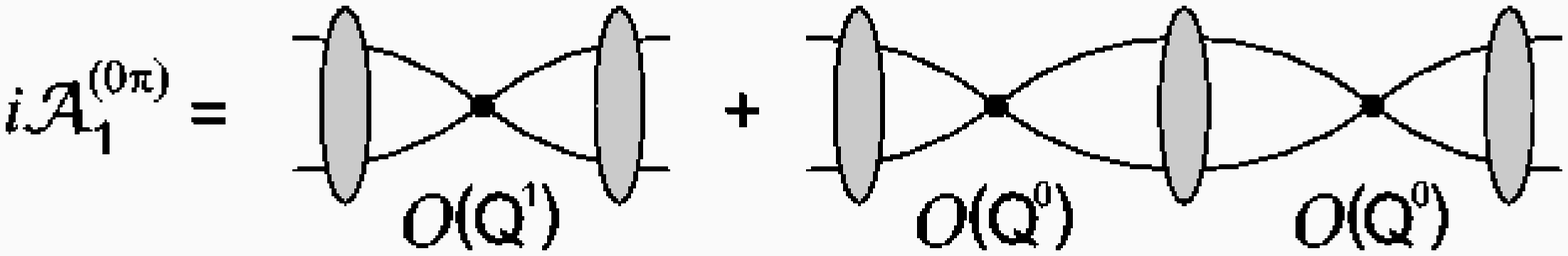}{.9}{Next-to-next-to-leading order contribution $\mcal{A}_{1}^{(0\pi)}$ from diagrams with no pion exchange.}
\beq
\mcal{A}^{(0 \pi)}_1 &=& {(1+ i\mcal{A}_{-1} L)}^2( - C_4 p^4 - B_{2,-1} p^2
- B_{0,1} )\nonumber\\
 &{  }& +i{(1+ i\mcal{A}_{-1} L)}^3 L {(- C_2 p^2 - B_{0,0})}^2.
 \eeq

\subsection{$\mcal{A}^{(\pi)}_1$}
\label{onepion}

All the diagrams of
Fig.~\ref{NNLO1pfig}  are generated by insertions of local operators
of order ${\cal O}(Q^0)$ and a single pion exchange. The one pion exchange contribution at this order is

\makefig{NNLO1pfig}{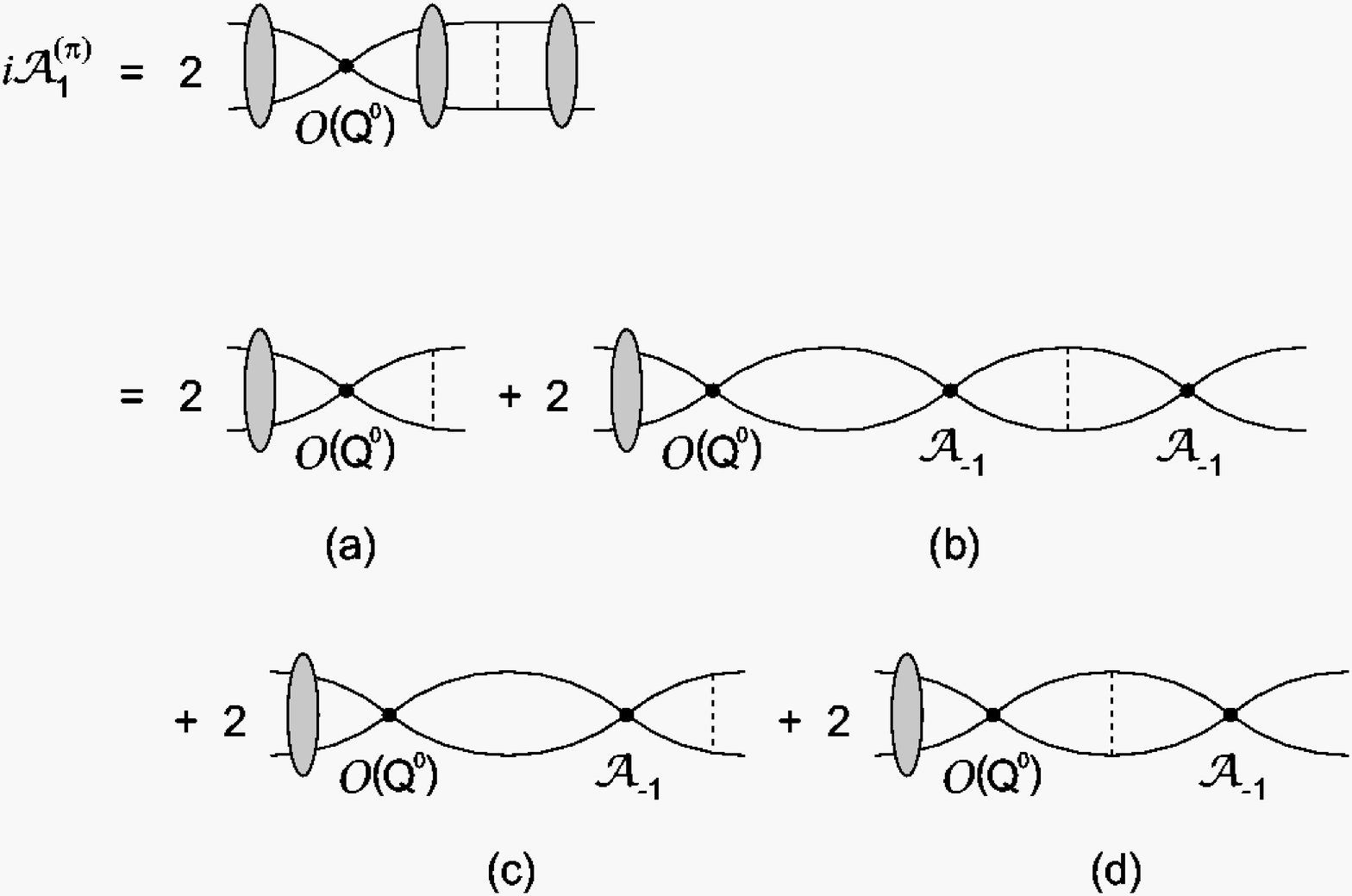}{3.2}{Next-to-next-to-leading order contribution $\mcal{A}_{1}^{(\pi)}$ from one pion exchange diagrams.}

\beq
\mcal{A}^{(\pi)}_1
&=&2 (1 + i
\mcal{A}_{-1}L)^2(-C_2 p^2 - B_{0,0})(P_1+P_2(i
\mcal{A}_{-1}))\nonumber\\
&{  }&+2 (1 + i \mcal{A}_{-1}L)\(-\frac{C_2}{2}\)(Q_1+Q_2(i\mcal{A}_{-1})),
 \eeq
where
\beq Q_1 &\equiv&\peta m^2_\pi M\(\frac{\mu}{2}\)^{4-D}\int\ddqm\frac 1 {({\bf q- p})^2 +
m^2_\pi}\nonumber \\ &\stackrel{PDS}\rightarrow&\peta m^2_\pi M
\(\frac{\mu -m_\pi}{4 \pi}\),
 \eeq
and
\beq Q_2 &\equiv&-i\peta m^2_\pi
M^2\(\frac{\mu}{2}\)^{8-2D}\int\ddqm\ddlm\frac 1 {{ l}^2 - { p}^2 -i\epsilon}\nonumber \\ 
&{}&\hspace{2.in}\times \frac 1{({\bf q- l})^2 + m^2_\pi}\nonumber\\
&=& Q_1 L\nonumber\\
&\stackrel{PDS}\rightarrow&-i\frac{g_A^2}{2 f^2}m_\pi^2 M^2 (\frac{\mu-m_\pi}{4 \pi})(\frac{\mu + i p}{4 \pi}),  \eeq
and $P_{1,2}$ are defined in Section~\ref{NLO}. In carrying out the subtractions in $Q_1$ and $Q_2$, some care is needed. The poles and the corresponding counterterms should be calculated treating $C_0$ perturbatively.

\subsection{$\mcal{A}^{(\pi \pi)}_1$}
\label{twopion}

\makefig{NNLO2pfig}{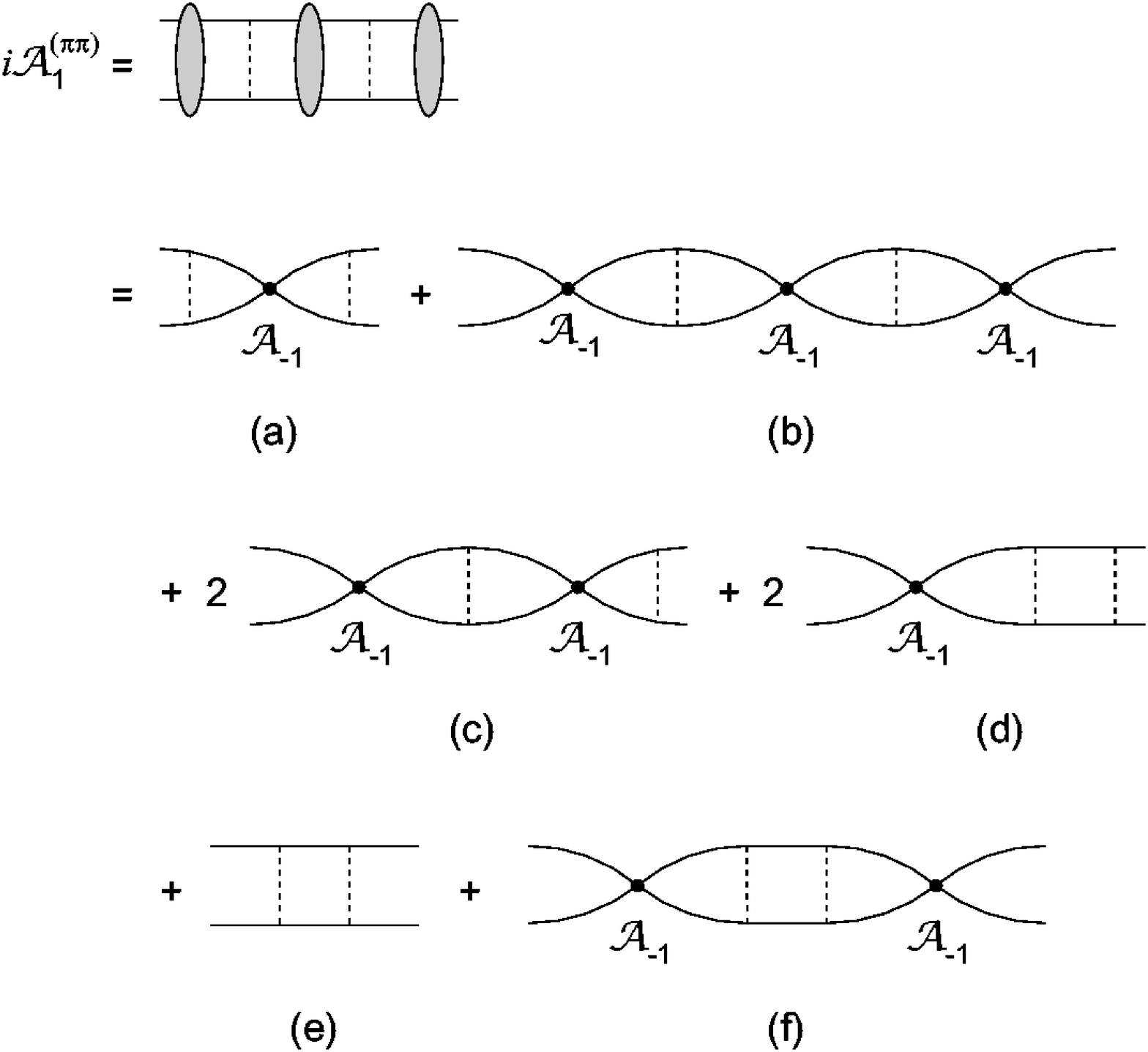}{4}{Next-to-next-to-leading order contribution $\mcal{A}_{1}^{(\pi\pi)}$ from two pion exchange diagrams.}

Fig.~\ref{NNLO2pfig} contains the diagrams of this
set. The graphs Fig.~\ref{NNLO2pfig}(e), (d) and (f) can be written in terms of
analytic functions $I_1(p)$, $I_2(p)$ and 
$I_3(p)$ respectively which are presented in the Appendix. The rest of the diagrams can be written in terms of
expressions that have already been defined. We find
\beq
\mcal{A}^{(\pi \pi)}_1
&=&-iI_1 +2(\mcal{A}_{-1})I_2
+i(\mcal{A}_{-1})^2 I_3\nonumber\\
& & + P_1(\mcal{A}_{-1})P_1 + 2 i(\mcal{A}_{-1})^2 P_1 P_2 - (\mcal{A}_{-1})^3(P_2)^2.
 \eeq

In the last two sections, we have calculated an expression for the LO$+$NLO $+$NNLO amplitude in terms of six independent free parameters $C_0$, $B_{0,0}$, $C_2$, $B_{0,1}$, $B_{2,-1}$ and $C_4$. The table below shows the new free parameters introduced at each order in the expansion.

\begin{center} 
\begin{tabular}{|l|c|c|c|}\hline
LO:& $C_{0}$& & \\ \hline
NLO:&$B_{0,0}$& $C_{2}$& \\ \hline
NNLO:&$B_{0,1}$& $B_{2,-1}$& $C_4$\\ \hline
\end{tabular}
\end{center}
\lskip
When the number of parameters is doubled (from NLO to NNLO) it is hard to consider an improvement of the fit as evidence for the convergence of the EFT expansion. Therefore, it is desirable to find theoretically viable ways to reduce the number of independent parameters. This is the subject of the next section.

\section{Determining the Effective Field Theory Couplings}
\label{Couplings}

\subsection{Matching conditions}
\label{Matching}

In this subsection I discuss a method for deriving conditions that reduce the number of free parameters. The guiding principle is simple: the leading couplings are defined in such a way that certain features of the low energy phase shift (like the effective range expansion parameters) are reproduced exactly. Conditions are then imposed on the sub-leading couplings so that these features are not modified at higher orders. Similar matching conditions were introduced in~\cite{mehen}.

From \Eq{pcot2A}, the
$Q$ expansion of $p\cot\delta$  in terms of the amplitude $\mcal{A}=\mcal{A}_{-1}+\mcal{A}_0+\mcal{A}_1+\cdots$, is
\beq
p\cot\delta=i p+ \frac{4\pi}{M \mcal{A}_{-1}}-\frac{4\pi}{M \mcal{A}_{-1}}\frac{\mcal{A}_0}{\mcal{A}_{-1}}+\frac{4\pi}{M \mcal{A}_{-1}}\left(\frac{\mcal{A}_0^2}{\mcal{A}_{-1}^2}-\frac{\mcal{A}_1}{\mcal{A}_{-1}}\right)+\cdots\ .
\eqn{Aexp}
\eeq
At low energies, the same quantity is given by
expansion  Eq.~\refeq{ERE2}:
\beq
p\cot\delta&=&-\gamma +\frac{1}{2}s_0\left (p^2+\gamma^2\right)+\frac{1}{2}s_1\left(p^2+\gamma^2\right)^{2}+\cdots\ .
\eqn{ERE2short}
\eeq
where $p\cot\delta$ has been expanded about $p^2=-\gamma^2$ (or $p=\pm i\gamma$).

The LO amplitude contributes only a constant to $p\cot\delta$. $C_0$ is determined by requiring
\beq
i p+ \frac{4\pi}{M \mcal{A}_{-1}}=-\gamma.
\eqn{C0eqn}
\eeq
This puts the pole in the amplitude at $p=i\gamma$.
At NLO and NNLO there are more constant contributions to $p\cot\delta$. The requirements that the position of the pole does not shift at NLO, nor at NNLO, reduce to:
\beq
{\rm NLO}:\hspace{.05in} \mcal A_0|_{p=-i\gamma}&=& 0\nonumber\\
{\rm NNLO}:\hspace{.05in} \mcal A_1|_{p=-i\gamma}&=& 0.
\eqn{nogammashift}
\eeq
These conditions determine the couplings $B_{0,0}$ and $B_{0,1}$. 
Similarly, one requires that $s_0$ in Eq.~\refeq{ERE2short} is determined at NLO by $C_2$:
\beq
\frac{\partial}{\partial p^2}\(-\frac{4\pi}{M \mcal{A}_{-1}}\frac{\mcal{A}_0}{\mcal{A}_{-1}}\)|_{p^2=-\gamma^2}=\frac{1}{2}s_0,
 \eeq
and that higher order (NNLO) effects do not change $s_0$:
\beq
{\rm NNLO}:\hspace{.05in} \frac{\partial}{\partial p} \mcal A_1|_{p=-i\gamma}&=& 0.
\eqn{nos0shift}
\eeq
This condition determines $B_{2,-1}$. (Eqns.~\refeq{nogammashift} and \refeq{nos0shift} can be more directly obtained by considering the expansion of the amplitude \Eq{EREinA}.) Note that at NNLO, fixing $\gamma$ is not exactly the same as fitting the scattering length $a$ whereas reproducing $s_0$ {\em is} essentially identical to fixing the effective range $r_0$.

The values of $\gamma$ and $s_0$ are used to determine $C_0$ and $C_2$. 
One can determine $C_4$ from $s_1$, the next free parameter in the expansion \Eq{ERE2short}. However, as in the theory without pions~\cite{ksw1,bira}, it can be shown using RG analysis that at NNLO, $C_4$ is completely determined. This is done below.

\subsection{Renormalization group flow}
\label{RG}

 The RG equations are derived by requiring that the
amplitude be exactly $\mu$ independent at each order in $Q$ for any
value of $p$.
The solutions to the RG equations for $C_{0}$ and $C_{2}$ are:
\beq \eqn{C0C2D2}
 C_{0}(\mu) &=& \frac{4\pi}{M}\(\frac{1}{\xi_{C_0}}-\mu\)^{-1}\nonumber\\
C_{2}(\mu) &=& \xi_{C_2} \(\frac{M C_{0}(\mu)}{4\pi}\)^2.
\eeq 

The integration constants $\xi_{C_0}$ and $\xi_{C_2}$  are determined by
boundary conditions.
The other coupling constants, denoted here by $X$, have the general solution
\beq
X(\mu)=\xi_X \(\frac{M C_{0}(\mu)}{4\pi}\)^2+f_{X}(C_{0}(\mu),B_{0,0}(\mu),C_{2}(\mu);\mu)
\eeq
where $f_X$ is known, and the only freedom in $X$ is in the constant $\xi_X$,
determined by boundary conditions.
Specifically,
\beq
B_{0,0}&=&\left(\frac{M C_{0}}{4 \pi}\right)^2 \left(\xi_{B_{0,0}}+\peta m_{\pi}^{2}\ln\left(\frac{\mu}{M}\right)\right)\nonumber\\
B_{0,1}&=&\frac{B_{0,0}^2}{C_{0}}-\peta m_\pi^2 \frac{M\mu}{4\pi}C_{2}+\xi_{B_{0,1}}\left(\frac{M C_{0}}{4\pi}\right)^2\nonumber\\
B_{2,-1}&=&\frac{2 B_{0,0} C_{2}}{C_{0}}+\xi_{B_{2,-1}}\left(\frac{M C_{0}}{4\pi}\right)^2\nonumber\\
C_4&=&\frac{C_{2}^{2}}{C_{0}}+\xi_{C_4}\left(\frac{M C_{0}}{4\pi}\right)^2.
\eqn{constraints}
\eeq

Examination of the solutions in \Eq{constraints} shows that the $\mu$ scaling of some couplings do not agree with their expected $Q$ counting. To determine the $Q$ counting one also needs to know the $Q$ counting of the RG constants $\xi_X$. That $\gamma$ can make up $Q$ counting for some of the $\xi_X$s was already shown in the theory without pions, where, for example, $C_{2,-1}\sim \gamma^2/(\mu-\gamma)^3$~\cite{gautam,chen1}. It is possible, however, to determine the $Q$ scaling of the leading couplings. We demonstrate this point by discussing directly the coupling $C_4$, which has power counting of $1/Q^3$. The second term in the solution to the RG equation for $C_4$ (\Eq{constraints}) is $\sim \xi_{C_4}/\mu^2$. In order for it to contribute at NNLO, $\xi_{C_4}$ should be $\mcal{O}(1/Q)$. Since $C_4$ has a sensible large scattering length ($\gamma\rightarrow 0$) limit, the possibility $\xi_{C_4}\sim 1/\gamma$ is ruled out. It is also assumed that $C_4$ is associated with short distance physics and thus it is fundamentally independent of pion physics. These considerations allow setting $\xi_{C_4}=0$ at this order, and have $C_4$ be completely determined by $C_0$ and $C_2$. There are in fact no new parameters at NNLO!

In the next section we check how well the EFT reproduces the phase
shift of the toy model and the real data when the matching conditions derived in the last two subsections are imposed.

\section{Results}
\label{REsults}
The ${}^1S_0$ scattering cross section $\sigma$ is related to the
phase shift and amplitude by

\beq\eqn{cross_section}
\sigma &=&\frac{4\pi}{p^2} \sin^2 \delta\nonumber\\
&=&\frac{M^2}{4\pi}|\mcal{A}|^2
\eeq
which can be expanded in powers of $Q$ ( writing $\sigma=\sigma_{-2}+\sigma_{-1}+\sigma_0+\cdots$):
\beq\eqn{sigmaA}
\sigma_{-2}&=&\frac{M^2}{4\pi}|\mcal{A}_{-1}|^2\nonumber\\
\sigma_{-1}&=&\frac{M^2}{4\pi}\left( \mcal{A}_0 \mcal{A}_{-1}^{*}+\mcal{A}_0^{*} \mcal{A}_{-1}\right)\nonumber\\
\sigma_0&=&\frac{M^2}{4\pi}\left(\mcal{A}_{-1}\mcal{A}_1^{*}+\mcal{A}_{-1}^{*}\mcal A_1+|\mcal{A}_0|^2\right).
\eeq 
 There are several advantages to using $\sigma$ as opposed to $\delta$ in comparing EFT result to data. The experimentally measured cross section $\sigma$ can be related to the sizes of the absolute value of the amplitude Eq.~\refeq{sigmaA}. $\delta$ is very sensitive to the value of $r_1$ and other higher moment coefficients and it is reflected in the value of $C_4$ that one chooses. This problem is avoided if one plots $\sigma$ or $\sin^2\delta\ $ instead, where the effects of $r_1$ or $\xi_{C_4}$ are truly sub-leading.  

In the following subsections we compare the EFT result with the exact phase shift from the toy model, and also do the corresponding comparisons for the real ${}^1S_0$ phase shift. We consider how well the data is reproduced by the EFT, whether the expansion is converging, and what is the breakdown scale.
  
\subsection{ Results for the toy model}
\label{toy_results}
For the toy model, $\gamma=-7.743\ {\rm MeV}$ and $s_0=1.584\ {\rm fm}$ and these values are used to determine the couplings. The results are presented in Fig~\ref{fig:toyfit}. Fig~\ref{fig:toyfit} (a) shows a clear improvement of fit as we go to higher order in the expansion. The NNLO result is a prediction as we do not introduce any new free parameter. Fig.~\ref{fig:toyfit} (b). shows the errors~\cite{error} at each order, $\Delta\equiv|\sin^2(\delta)_{EFT}-\sin^2(\delta)_{exact}|$, on a log-log scale. The breakdown scale, where the errors become comparable, can be read off of the error plot to be $\sim 520$ MeV which is slightly higher than the expected $m_\rho/2$.
 
\maketwofigs{fig:toyfit}{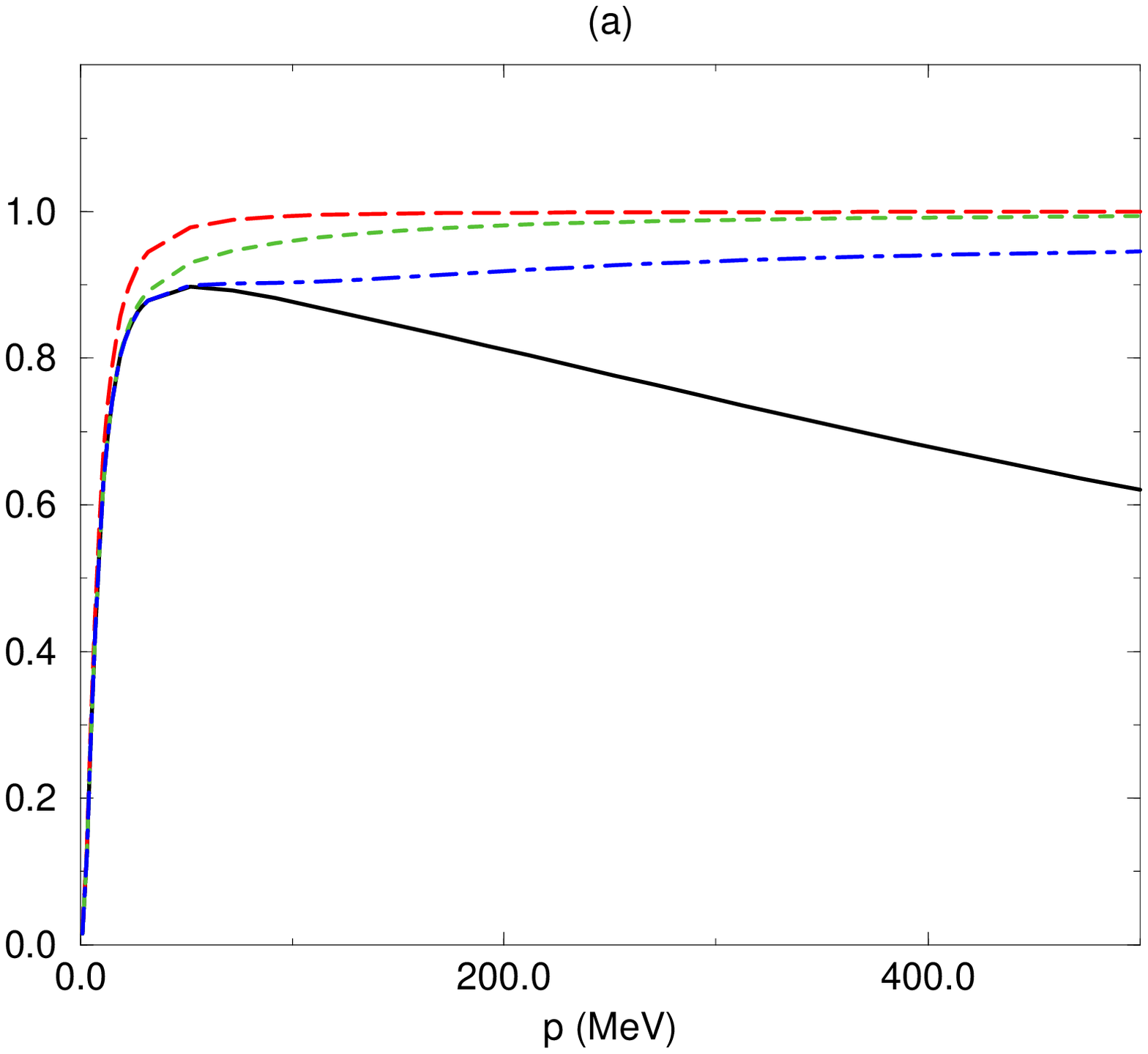}{-0.32}{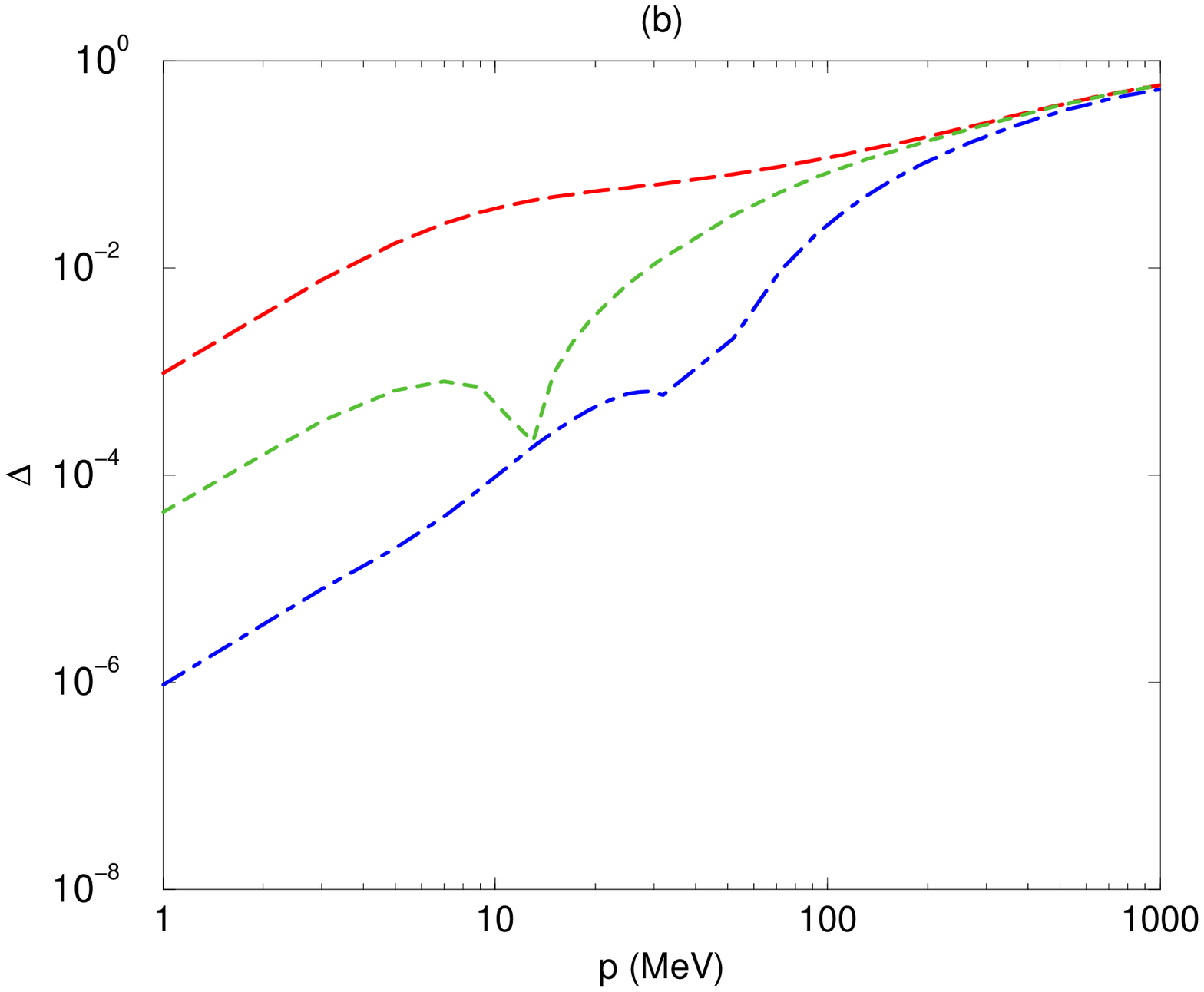}{2.2}{\protect{Toy model: (a) $\sin^2(\delta)$ vs. $p$(MeV); (b) log-log plot of $\Delta\equiv  |\sin^2(\delta)_{EFT}-\sin^2(\delta)_{exact}|$ vs. $p$(MeV). The exact phase shift is described by the solid curve. The long-dashed curves denote the LO phase shift and error, the dashed curves are the NLO results, and the dot-dashed curves describe the NNLO results.}}


Table~\ref{table:toy} shows the couplings at $\mu=m_\pi$. The entries are given
in units of ${\rm fm}^2$. To make a direct comparison of the
  couplings easier, we included the momentum factors from the operators
  $C_2 p^2,\ B_{2,-1} p^2$ and $C_4 p^4$, evaluated at $p=m_\pi$. From the table we see a hierarchy in the sizes of the couplings corresponding to the
orders in the expansion. This is a verification of the power counting scheme. A rough estimate of the expansion parameter is $Q/\Lambda\sim m_\pi/(m_\rho/2)\sim1/3$, which is also suggested by the table.\vspace{.2 in}
\begin{table}[t]\caption{Renormalized couplings for the toy model at $\mu=m_\pi$ in units of ${\rm fm}^2$.\label{table:toy}}
\begin{center} 
\begin{tabular}{|c||c|c||c|c|c|}\hline
\multicolumn{1}{|c||}{LO}&\multicolumn{2}{|c||}{NLO}&\multicolumn{3}{|c|}{NNLO}\\ \hline
  $C_{0}$ & $C_{2} m_\pi^2$ & $B_{0,0}$ & $B_{0,1}$ & $\ B_{2,-1}\ m_\pi^2\ $ & $C_4\ m_\pi^4$\\ 
\hline $\  -3.526\ $ & $\ 0.06912\ $ &$\ 1.412\ $ & $\ 0.5620\ $ & $\ 0.2886\ $ & $\ -0.001355\ $\\
\hline
\end{tabular}
\end{center}
\end{table}

\subsection{Results for the real ${}^1S_0$ scattering}
\label{Nijmegen}

Let us recall that in the KSW power counting the LO amplitude, which is
inversely proportional to the momentum $p$, is calculated in terms of
a non-perturbative dimension 6 four-nucleon operator. The other higher
derivative operators and pion exchanges are treated
perturbatively. The expansion parameter is $Q/\Lambda$ with $p$,
$m_\pi\sim Q$ and $\Lambda$ some high energy cut-off, and at $p\sim Q$ the LO amplitude scales as
$1/Q$. Pion production, however, at external momentum $p=\sqrt{m_\pi
  M}\sim 360\ {\rm MeV}$, introduces a new scale, and it is convenient
to use a different expansion parameter,
$Q_r\sim\sqrt{m_\pi M}$, instead of $Q\sim Q_r^2/M\ll
Q_r$~\cite{mehen}. At $p\sim Q_r$, the LO amplitude scales as $1/Q_r$. The
relative importance of the potential, radiation and soft pion
contributions depends on the physics associated with different scales determined by the momentum $p$. 

 At $p\sim Q_r$, potential and soft pions contribute at $Q_r^0$ and $Q_r^2$ respectively, while  radiation pions contribute at $Q_r^3$~\cite{mehen}. Thus, for an NNLO ($\mcal{O}(Q_r)$) calculation, only potential pions contribute at this scale. At $p\sim Q$, potential pions contribute at order $Q^0$. The power counting for the soft and radiation pion is less clear. Examination of the radiation pion diagrams that have been calculated in ~\cite{mehen}, shows that some diagrams, which scale as $Q_r^3$ at high momentum, become order $Q^{1/2}$ at low external momentum. Such contributions seem to violate the presumed expansion of the amplitude in integer powers of Q, but in fact there is a cancellation between the different diagrams, and the leading contribution from radiation pion graphs starts at $\mcal{O}(Q)$. Furthermore, the remaining terms are highly suppressed because of Wigner's $SU(4)$ spin-isospin symmetry~\cite{mehen2} and can be neglected in the NNLO calculation (see Tom Mehen's discussion in these proceedings).
We do not see (at the moment) a general reason for such cancellations (of order $Q^{1/2}$), and it is in principal possible that radiation or soft pion graphs that are of high order in $Q_r$ at high momentum, become important at low momentum, and should in fact be included in an NNLO calculation of the amplitude. Here, however, this option will not be investigated, and only the potential pion contributions will be taken into account, admittedly with the possibility of having left out some graphs. The EFT amplitude for the toy model with potential pions is identical to the potential pion contribution to the real ${}^1 S_0$ amplitude, after a redefinition of $C_{0,0}\rightarrow C_{0,0}-\peta$. With these considerations, the EFT amplitude with only potential pions is applied to the real ${}^1S_0$ scattering. We use $\gamma=-7.889\ {\rm MeV}$ ($a=-23.68\ {\rm fm}$) and $s_0=2.73\ {\rm fm}$ to fix the couplings.

Here again, we have a prediction at NNLO, Fig.~\ref{fig:realfit}. By going to higher order we get better result, Fig.~\ref{fig:realfit}(a). From Fig.~\ref{fig:realfit}(b), the breakdown scale is estimated to be $\sim 400$MeV. Just like in the toy model, a rough estimate of the breakdown scale is $m_\rho/2$ due to the exchange of vector meson $\rho$. 

\maketwofigs{fig:realfit}{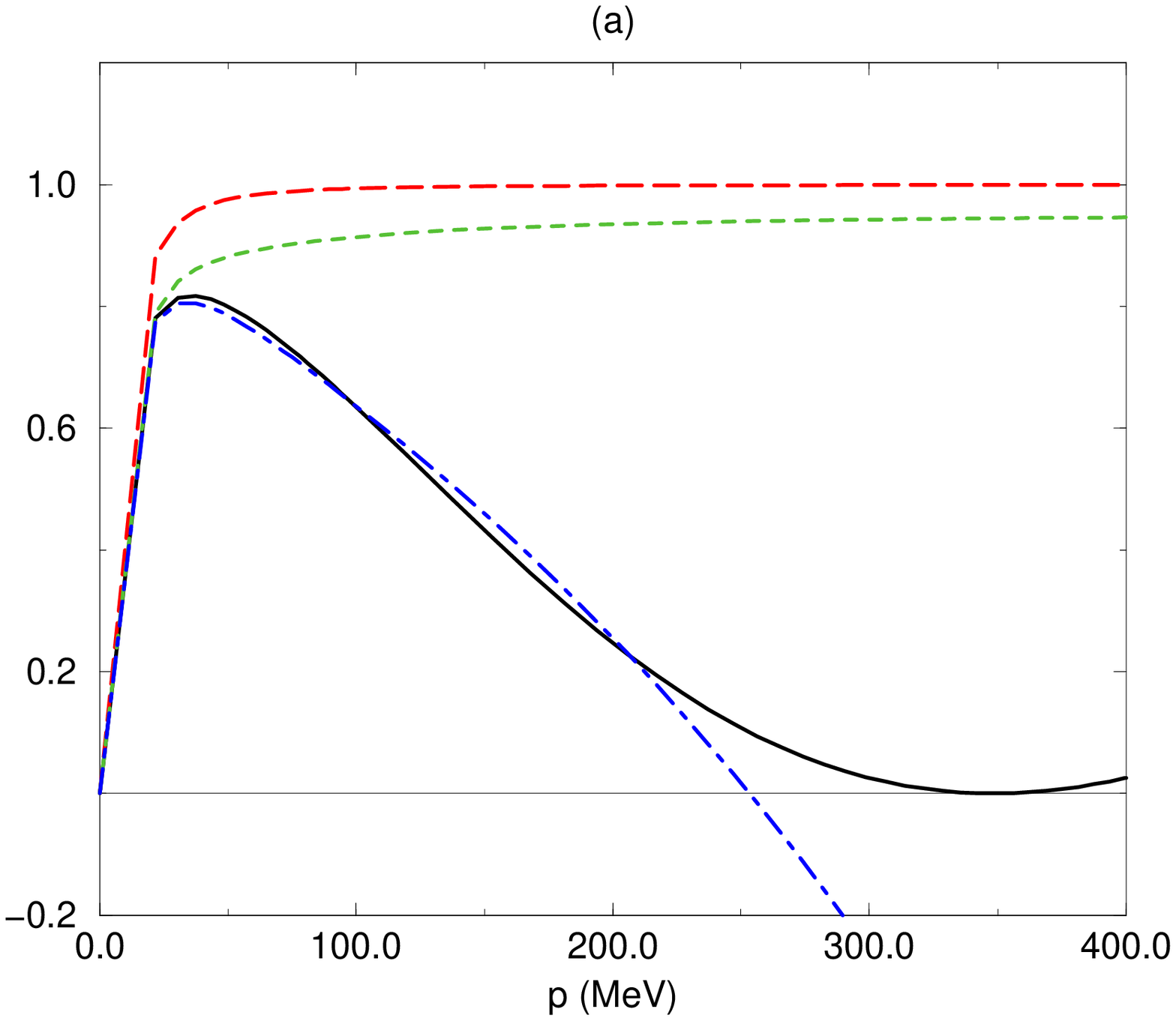}{-0.32}{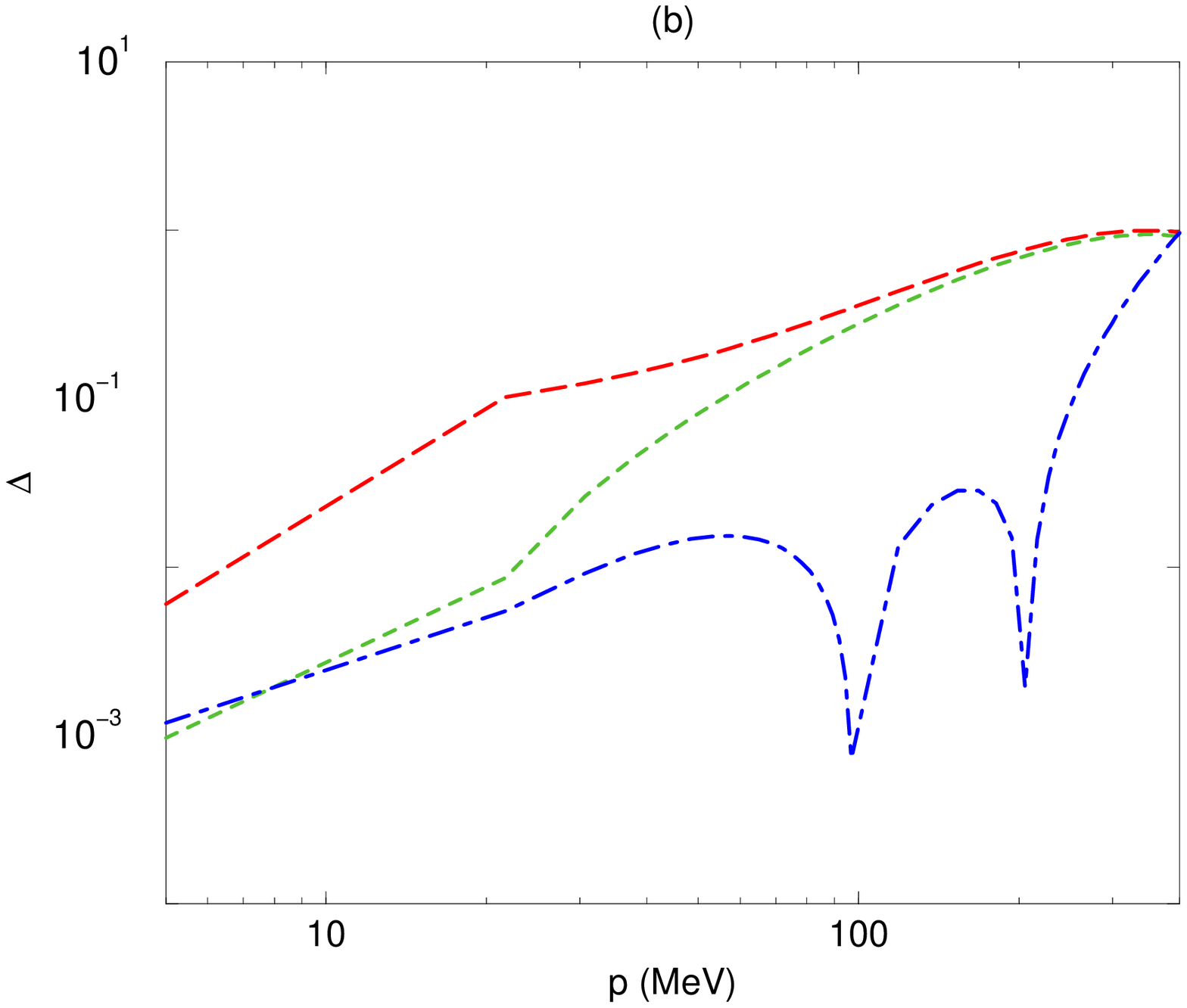}{2.2}{\protect {Nijmegen: (a) $\sin^2(\delta)$ vs. $p$(MeV); (b) log-log plot of $\Delta\equiv 
|\sin^2(\delta)_{EFT}-\sin^2(\delta)_{Nijmegen}|$ vs. $p$(MeV). The Nijmegen phase shift is described by the solid curve. The long-dashed curves denote the LO phase shift and error, the dashed curves are the NLO results, and the dot-dashed curves describe the NNLO results.}}

Table~\ref{table:real} shows the couplings at $\mu=m_\pi$. The entries are given
in units of ${\rm fm}^2$. We included the momentum factors from the operators
  $C_2 p^2,\ B_{2,-1} p^2$ and $C_4 p^4$, evaluated at $p=m_\pi$. The sizes of the couplings at various order in the expansion suggest convergence of the expansion with $Q/\Lambda\sim 1/3$.\vspace{.2 in}
\begin{table}[t]\caption{Renormalized couplings for the real ${}^1S_0$ channel at $\mu=m_\pi$ in units of ${\rm fm}^2$.\label{table:real}}
\begin{center} 
\begin{tabular}{|c||c|c||c|c|c|}\hline
\multicolumn{1}{|c||}{LO}&\multicolumn{2}{|c||}{NLO}&\multicolumn{3}{|c|}{NNLO}\\ \hline
  $C_{0}$ & $C_{2} m_\pi^2$ & $B_{0,0}$ & $B_{0,1}$ & $\ B_{2,-1}\ m_\pi^2\ $ & $C_4\ m_\pi^4$\\ 
\hline $\ -3.5226\ $ & $\ 1.39685\ $ &$\ 1.41738\ $ & $\ 0.56156\ $ & $\ 0.541789\ $ & $\ -0.553904\ $\\
\hline
\end{tabular}
\end{center}
\end{table}

\section{Summary and Conclusion}
\label{Conclusions}

A NNLO calculation for nucleon-nucleon scattering in the ${}^1S_0$ channel  using potential pions was presented. Radiation and soft pions do not contribute at this order for center of mass momentum $p\sim Q_r$. However, there could be soft and radiation pion contributions at momentum $p\sim Q$ which have not been calculated yet. If for low momentum $p\sim Q$, soft and radiation pion contributions are found to be small at NNLO, then this calculation gives the NNLO nucleon-nucleon scattering amplitude in the ${}^1S_0$ channel for momentum $p\leq m_\rho/2$. 

The calculation was applied to a toy model and the real data. The toy model provided a setup where only potential pions contribute by construction. This allowed us to address some general questions regarding the convergence of KSW power counting, number of free parameters and the role of renormalization group flow, without worrying about unaccounted for radiation and soft pion contributions. 

At NNLO there are six independent parameters. In the effective field theory, these parameters are associated with contact operators which encodes information about the high energy physics. However, in the singlet channel the high energy physics conspire to produce a low energy scale  $\gamma\sim 8$ MeV. This small energy scale has to be taken into account in using dimensional analysis to determine the sizes of the parameter. RG analysis, which only determines the $\mu$ scaling of parameters, is not always enough to determine their $Q$ scaling. Some of the parameters such as $\xi_{C_4}$ can be determined from RG analysis. We introduce a method to determine the rest of the parameters. This is done by matching the EFT amplitude at low momentum to the effective range expansion and all the parameters are given in terms of two experimental inputs, $\gamma$ and $s_0$. We find that there are no free parameters at NNLO and we make a prediction for the high momentum behavior of the scattering amplitude.

For the toy model we find that including higher order terms in the amplitude improves the fit to ``data''. The hierarchy in the sizes of couplings show converges with an expansion parameter of $Q/\Lambda\sim 1/3$. The breakdown scale is estimated to be $\sim 520$ MeV by comparing the sizes of errors at different order. This is slightly higher than the expected scale $m_\rho/2$.

The fit to real data shows significant improvement at NNLO. We see an hierarchy in the sizes of the couplings which suggests an expansion parameter similar to the toy model $Q/\Lambda\sim 1/3$. The breakdown scale is estimated to be $\sim 400$ MeV. 

\section*{Acknowledgments}

I thank Noam Shoresh for a rewarding collaboration. I would also like to thank Paulo Bedaque, Harald Grie{\ss}hammer, David Kaplan, Tom Mehen, Dan Phillips, Martin Savage, James Steele and Iain Stewart for many useful discussions. It is also a pleasure to thank the organizers of the conference and the participants. This work is supported in part by U.S. Department of Energy Grant DE-FG03-97ER41014.

\section*{Appendix: The integrals $I_1$, $I_2$ and $I_3$}


Fig.~\ref{NNLO2pfig} (e) is given by:

\beq
I_1&=&i\(\frac{{g^2}}{2{f^2}}
  \)^2 m_{\pi }^{4} M \bar{I}_1\nonumber\\
\bar{I}_1&=&\left<\int\dtqm \frac{1}{q^2-p^2-i\epsilon}\frac{1}{(\bfq-\bfp)^2+m_\pi^2}\frac{1}{(\bfq-\bfpp)^2+m_\pi^2}\right>\nonumber\\
&=&\frac{1}{32\pi m_\pi^3 \rho^3}\left(i L(\rho)+ M(\rho)\right)\\
L(\rho)&=&\int\limits_0^{1/2} dt \frac{1}{t(1-t)} \ln\left(\frac{(1+4\rho^2 t )(1+4\rho^2(1-t))}{1+4\rho^2}\right)\nonumber\\
&=&{\frac{1}{2}{{\ln^2 (1 + 4\,{{\rho }^2})}}}\nonumber\\
M(\rho)&=&2\int\limits_0^{1/2} dt \frac{1}{t(1-t)}\left(\tan^{-1}(2\rho\sqrt{1+4\rho^2t(1-t)})-\tan^{-1}(2\rho)\right)\nonumber\\
&=&{\frac{i}{2}}\,{{\pi }^2} - i\,{{\ln^2 \(2\)}} + 
  {\frac{i}{2}}\,\ln \(4\)\,\ln \(1 - 2\,i\,\rho \) + 
  2\,\pi \,\ln \(2 - 2\,i\,\rho \)\nonumber\\
& &- i\,\ln \(2\,{{\( 1 + i\,\rho  \) }^2}\,
     \( 1 - 2\,i\,\rho  \) \)\,
   \ln \(1 + 2\,i\,\rho \)\nonumber\\
&{}& -\tan^{-1} \(2\,\rho \)\,\ln \({\frac{4\,
        {{\( 1 + {{\rho }^2} \) }^2}}{1 + 
        4\,{{\rho }^2}}}\)-   2\,\tan^{-1} \(\rho \)\,\ln \(1 + 4\,{{\rho }^2}\) \nonumber\\
& & 
 - 
  2\,i\,Li_2\(
    {\frac{1}{2}} + i\,\rho \) - 2\,i\,Li_2\(-1 - 2\,i\,\rho \) - 2\,i\,Li_2\(2 - 2\,i\,\rho \)\nonumber,
\eeq 
where the average is over the outgoing momentum $\bfpp$. $\rho =p/m_\pi$ and the dilogarithm, $Li_2(z)$, is defined as
\beq
Li_2(z)=-\int\limits_0^{z} dt \frac{\ln(1-t)}{t}\nonumber.
\eeq

Fig.~\ref{NNLO2pfig} (d) is given by:

\beq
I_2&=&\(\frac{{g^2}}{2{f^2}}\)^2 m_{\pi }^{4} {M^2}\bar{I}_2\nonumber\\
\\
\bar{I}_2&=&\int\dtqm\dtlm \frac{1}{q^2-p^2-i\epsilon}\frac{1}{(\bfq-\bfp)^2+m_\pi^2}\frac{1}{l^2-p^2-i\epsilon}\frac{1}{(\bfq-\bfl)^2+m_\pi^2}\nonumber\\
&=&\frac{i}{128 \pi^3 m_{\pi}^2\rho}\int_{-\infty}^{\infty}dx{\ln\(\frac{i+x+\rho}{i-x+\rho}\)\ln\(\frac{(x+\rho)^2+1}{(x-\rho)^2+1}\)}\frac{1}{x^2-\rho^2-i\epsilon}\nonumber\\
&=&
\frac{1}{128\,
     {{\pi }^2}\,{{\rho }^2}\,{{{m_{\pi }}}^2}}
\Bigl(-{{\pi }^2} + 4\,i\,\pi \,
      \ln \(2 - 2\,i\,\rho \) + 
     \ln \(2\)\,\ln \(1 + 2\,i\,\rho \) 
\Bigr. \nonumber \\ & &\hspace{.5 in} \left.
+ 
     \ln \(1 - 2\,i\,\rho \)\,
      \left[4\ln \( 1 - i\,\rho\)-\ln\({\( 1 - 2\,i\,\rho \)\( 1 + 4\,\rho^2\)}/{8}\) \right]
\right. \nonumber \\ & &\hspace{.7 in} \left.
- Li_2\({\frac{1}{2}} - i\,\rho \) 
+ Li_2\({\frac{1}{2}} + i\,\rho \) + 
     4\,Li_2\(2 - 2\,i\,\rho \)\).\nonumber
\eeq

From Fig.~\ref{NNLO2pfig}(f), we define
\beq
I_3&=&-i \(\frac{{g^2}}{2{f^2}}\)^2 m_{\pi }^{4} {M^3} \bar{I}_3\nonumber\\
\bar{I}_3&=& \int\dtkm\frac{1}{k^2-p^2-i\epsilon}I(\bfk,\bfp)^2\nonumber\\
I(\bfk,\bfp)&=&\int\dtqm\frac{1}{q^2-p^2-i\epsilon}\frac{1}{(\bfq-\bfk)^2+m_\pi^2}.
 \eeq
and it gives
\beq
\bar{I}_3&=&\frac{i}{64\pi^3 p}\int\limits_0^{1/2} dt\frac{1}{t(1-t)}\int\limits_0^\infty ds e^{-(m-2i p)s}\frac{e^{-i 2 p s t}-1}{s}\nonumber\\
&=& \frac{i}{64 \pi^3 \rho m_\pi}\left( \ln\left(\frac{1-i\rho}{1-i 2\rho}\right)\ln(i\rho)+Li_2\left(\frac{-i\rho}{1-i2\rho}\right)\right.\nonumber\\
&{}&+\Bigl. Li_2(1-i\rho)-Li_2(1-i2\rho)\Bigr).
\eeq

The explicit expressions for $I_1$, $I_2$ and $I_3$ presented above are valid only for $\rho \geq 0$ and small positive imaginary $\rho$ ($|\rho|\ll 1$). For other values of $\rho$, the integrals have to be done taking into account the poles and cuts in the integrand appropriately.

\end{document}